\documentclass[reprint, amsmath,amssymb, aps, longbibliography]{revtex4-1}
\usepackage{geometry}                
\usepackage{graphicx}
\usepackage{array} 
\usepackage{dcolumn}
\usepackage{bm}
\usepackage{color} 
\usepackage{hyperref}
\usepackage{ulem}

\begin{document}

\title{Arm-Locking with the GRACE Follow-On Laser Ranging Interferometer}

\author{James Ira Thorpe}
\email{james.i.thorpe@nasa.gov}
\affiliation{Gravitational Astrophysics Laboratory, NASA Goddard Space Flight Center, Greenbelt, MD 20771, USA}

\author{Kirk McKenzie}
\affiliation{Jet Propulsion Laboratory, California Institute of Technology, Pasadena, California 91109, USA}

\date{\today}

\begin{abstract}
Arm-locking is a technique for stabilizing the frequency of a laser in an inter-spacecraft interferometer by using the spacecraft separation as the frequency reference. A candidate technique for future space-based gravitational wave detectors such as the Laser Interferometer Space Antenna (LISA), arm-locking has been extensive studied in this context through analytic models, time-domain simulations, and hardware-in-the-loop laboratory demonstrations. In this paper we show the Laser Ranging Interferometer instrument flying aboard the upcoming Gravity Recovery and Climate Experiment Follow-On (GRACE-FO) mission provides an appropriate platform for an on-orbit demonstration of the arm-locking technique. We describe an arm-locking controller design for the GRACE-FO system and a series of time-domain simulations that demonstrate its feasibility.  We conclude that it is possible to achieve laser frequency noise suppression of roughly two orders of magnitude around a Fourier frequency of 1Hz with conservative margins on the system's stability. We further demonstrate that `pulling' of the master laser frequency due to fluctuating Doppler shifts and lock acquisition transients is less than $100\,$MHz over several GRACE-FO orbits. These findings motivate further study of the implementation of such a demonstration.

\end{abstract}


\maketitle

\section{Introduction}
\label{sec:intro}
Space-based interferometric gravitational wave instruments such as the Laser Interferometer Space Antenna (LISA)\cite{Bender_98, ESA_Yellow_Book} sense fluctuations in spacetime curvature by measuring the distance between freely-falling test masses over large baselines using heterodyne interferometry. The three LISA spacecraft are placed in independent orbits that produce a triangular constellation that is approximately equilateral but experiences distortions at the $\sim 1\%$ level over the lifetime of the mission\cite{Hughes_08}. The resulting unequal baselines provide a pathway for laser frequency noise to couple into the displacement measurement. To compensate for this, LISA employs Time Delay Interferometry (TDI) \cite{Armstrong_99}, a technique which takes advantage of the fact that the accumulated phase in each baseline is independently measured and allows the synthesis of an effective equal-arm interferometer in ground-based post-processing \cite{Shaddock_04b}.  TDI's ability to suppress laser frequency noise is limited by knowledge of the constellation baselines, resulting in a requirement on the residual laser frequency noise in the primary LISA laser of $\sim 300\,\mbox{Hz}/\sqrt{\mbox{Hz}}$ in the LISA measurement band, $0.1\,\mbox{mHz}\leq f \leq 100\,\mbox{Hz}$\cite{Thorpe10}. This is some four orders of magnitude lower than the `free-running' noise performance of candidate laser systems, which have a residual noise level of roughly $30\,\mbox{kHz}/\sqrt{\mbox{Hz}}\cdot\left(1\,\mbox{Hz}/f\right)$. As a result, LISA-like missions require some form of active frequency stabilization of their primary light sources.

Arm-locking \cite{Sheard_03, Sutton_08} is a technique developed for LISA that utilizes one or more of the constellation baselines as a frequency reference. Arm-locking exploits the mismatch of the constellation arms (or the mismatch between one constellation arm and a short path on an optical bench in the case of `single-arm' arm-locking) to measure and subsequently suppress phase noise. Arm-locking in the LISA context has been thoroughly studied using both analytic \cite{McKenzie_09} and numeric \cite{Thorpe11} techniques, with particular emphasis placed on implementation details such as lock acquisition and `frequency pulling' of the arm-locked laser.

The Gravity Recovery And Climate Explorer Follow-On (GRACE-FO) mission is a collaboration between NASA and the German Geosciences Research Center to measure time-variations in the geoid, Earth's gravitational field. GRACE-FO is a successor to the GRACE mission\cite{Tapley04}, which has been performing geodetic measurements since 2002. GRACE and GRACE-FO employ the same basic measurement scheme: two spacecraft flying in a common low-Earth orbit with one following $170\,$km to $270\,$km behind the other. Variations in the underlying geoid modulate this range, which is measured using a microwave ranging system\cite{Dunn2003} and corrected for atmospheric drag and other effects using data from on-board accelerometers and GPS\cite{Case2010}.  GRACE-FO, which is expected to launch in 2017, will include a parallel optical ranging system known as the Laser Ranging Interferometer (LRI)\cite{Sheard12}, which shares much of its system design with the interferometric measurement system developed for the LISA mission.

In this paper we describe how the LRI on GRACE-FO could be used to perform an on-orbit demonstration of arm-locking, further increasing technical readiness and reducing risk for employing this technology in LISA. A similar demonstration of the TDI technique using the GRACE-FO LRI\cite{Francis15} could be conducted as part of the same program. In section \ref{sec:design} of this paper, we compare the relevant parameters of the GRACE-FO and LISA systems and present our design of an arm-locking controller for GRACE-FO. In section \ref{sec:methods}, we describe a series of time-domain simulations that were used to evaluate this candidate design in terms of noise performance and impact on other system elements, such as the laser. In section \ref{sec:results}, we present the results of the simulations which are further discussed in section \ref{sec:discuss}.

\section{Design}
\label{sec:design}
\subsection{Arm-locking for LISA and GRACE-FO}
Arm-locking is a technique to stabilize the frequency of a light source to a length reference provided by an optical delay line or arm. In both the LISA and GRACE-FO  cases, one optical bench on one of the spacecraft is designated as the master and the light from its laser is transmitted to the distant spacecraft. The far spacecraft operates in `repeater' mode; measuring the phase of the incoming light field relative to that of its local laser source and adjusting the local laser to match it via a high-gain phase-lock loop. The light from the far spacecraft then travels back to the master spacecraft, where it is interfered with a beam from the master laser.  The relationship between the master laser frequency and the frequency measured at this interference is characterized by the arm transfer function:
\begin{equation}
T_{arm}(f) \equiv 1 - \exp\left(-2\pi i f \tau\right),\label{eq:Tarm}
\end{equation}
where $\tau$ is the round-trip light travel time through the arm. For Fourier frequencies $ f \ll 1/\tau$, $T_{arm}\approx 2\pi i f \tau$, which is the transfer function of a first derivative (scaled by a constant factor $\tau$). In this regime, it is relatively straightforward to design a controller, say $G(f)\propto f^{-2}$, that can take this estimate of the derivative of laser frequency noise and use it to stabilize the noise of the master laser. 

However, in the case of LISA, $\tau \sim 33\,\mbox{s}$, meaning that the bandwidth of this type of arm-locking controller would be limited to $1/33\,\mbox{s}\approx 30\,\mbox{mHz}$, which lies in the LISA science band.  Employing a more sophisticated scheme using two of the three arms in the LISA constellation can effectively transform $\tau$ to the \textit{difference} in the round-trip times between the two arms\cite{Sutton_08}, which is typically $\sim 500\,\mbox{ms}$. This would permit a bandwidth of $\sim 2\,$Hz, still too small to allow for any significant gain in the science band.

To extend the bandwidth of the arm-locking system to frequencies greater than $1/\tau$, the controller must account for the phase response of $T_{arm}(f)$ near the `null' frequencies $f_n \equiv n/\tau,\:n=1,2,3\ldots$. As $f\rightarrow f_n$ from below, the phase drops towards approaches $-\pi/2$ and then jumps to $+\pi/2$ just above $f_n$. To provide phase margin near $f_n$, the controller must have a roll-off of less than unity, $G(f)\propto f^{-\alpha},\:0\leq\alpha\leq 1$ for $ f \geq 1/\tau$. Systems with controllers of this design have been demonstrated to be stable in analytical\cite{Sutton_08, McKenzie_09}, numerical\cite{Thorpe11}, and hardware\cite{Wand_09} models with several hundred nulls included in the controller bandwidth.

A second challenge with arm-locking results from the response of $T_{arm}(f)$ for low frequencies, which goes to zero as $f\rightarrow 0$.  This means that any low-frequency or constant offsets in the arm-locking error signal will overwhelm the signal from the residual laser noise and can cause the frequency of the master laser to be `pulled'.  For example, the arm-locking error signal in both LISA and GRACE-FO will contain an offset due to the round-trip Doppler shift caused by the relative motion between the spacecraft. For the case of a constant offset and an arm-locking controller with large DC gain, the master laser will be pulled at a rate:

\begin{equation}
\left.\frac{d\nu}{dt}\right|_{G(0)\rightarrow\infty}\approx \frac{\Delta}{\tau},\label{eq:dopPull}
\end{equation}
where $\nu(t)$ is the frequency of the master laser and $\Delta$ is the constant round-trip Doppler. For single-arm arm-locking in LISA, this would cause a pulling rate of $20\,\mbox{MHz}/33\,\mbox{s}\approx 600\,\mbox{kHz}/\mbox{s}$ meaning that the laser would be pulled through a single-mode region, typically several GHz wide, in roughly three hours. Two ways to mitigate this frequency pulling are (i), estimate and subtract the Doppler signal in a feed-forward scheme; and (ii), reduce the frequency pulling by reducing the gain of the controller below the science band, an approach colloquially referred to as `AC coupling'. Arm-locking system designs for LISA employing both of these techniques have been successfully demonstrated analytically\cite{McKenzie_09} and numerically\cite{Thorpe11}.

\begin{table}[h!]
\caption{\label{tab:LISAvGRACE}Comparison of key parameters for arm-locking in LISA and GRACE-FO. For parameters noted $\mbox{with}\:^\dagger$, the values in parentheses refer to typical \textit{differences} between pairs of LISA arms that are used in Dual Arm-Locking schemes.}
\begin{center}
\begin{tabular}{|c|c|c|}
\hline
Parameter & LISA & GRACE-FO \\
\hline
\hline
Baseline$^\dagger$ & 
{$\!\begin{aligned}
               5\times 10^6\,\mbox{km} \\    
               (8\times 10^4\,\mbox{km}) \end{aligned}$} 
& 
{$\!\begin{aligned}
               170\,\mbox{km}\:min\\    
               270\,\mbox{km}\:max\end{aligned}$}  \\
\hline
Round-trip delay$^\dagger$ & 
{$\!\begin{aligned}
               33\,\mbox{s} \\    
               (500\,\mbox{ms}) \end{aligned}$}                
& $1.8\,$ms \\
\hline
Transmitted power & 2 W & 25mW \\
\hline
Aperture & 30 cm & 1 cm \\
\hline
Received power & 100 pW & 100 pW \\
\hline
Doppler Amplitude & 20 MHz & $\gtrsim 4\,$MHz \\
\hline
Dopper Period & 1 year & 90 min \\
\hline
\end{tabular}
\end{center}
\end{table}%

Table \ref{tab:LISAvGRACE} shows a comparison between LISA and GRACE-FO of the key parameters for an arm-locking system. Compared with a dual arm-locking scheme for LISA, GRACE-FO has a round-trip delay that is nearly three hundred times shorter.  While the telescope aperture, baseline, and transmitted laser power numbers are quite different for the two systems, the resulting received light power levels are similar, meaning that the level of photon shot noise that limits laser frequency measurements is similar.  This is not a coincidence but rather a result of the fact that the LRI system is based on hardware (e.g. photoreceivers, phase meters, etc.) developed for LISA.  

The magnitude of the Doppler signals for LISA and GRACE-FO are similar, although the period of the fundamental variation is $\sim 6\times 10^3$ times shorter for GRACE-FO, meaning that the Doppler derivative is much larger.  Also, the GRACE-FO Doppler signal will contain significant contributions from the effect of the geoid, the main science signal, in the frequency band of interest. This is discussed in more detail in section \ref{sec:link}.

\subsection{GRACE-FO Arm-locking System Design}
\begin{figure}[h]
\begin{center}
\includegraphics[width=8 cm]{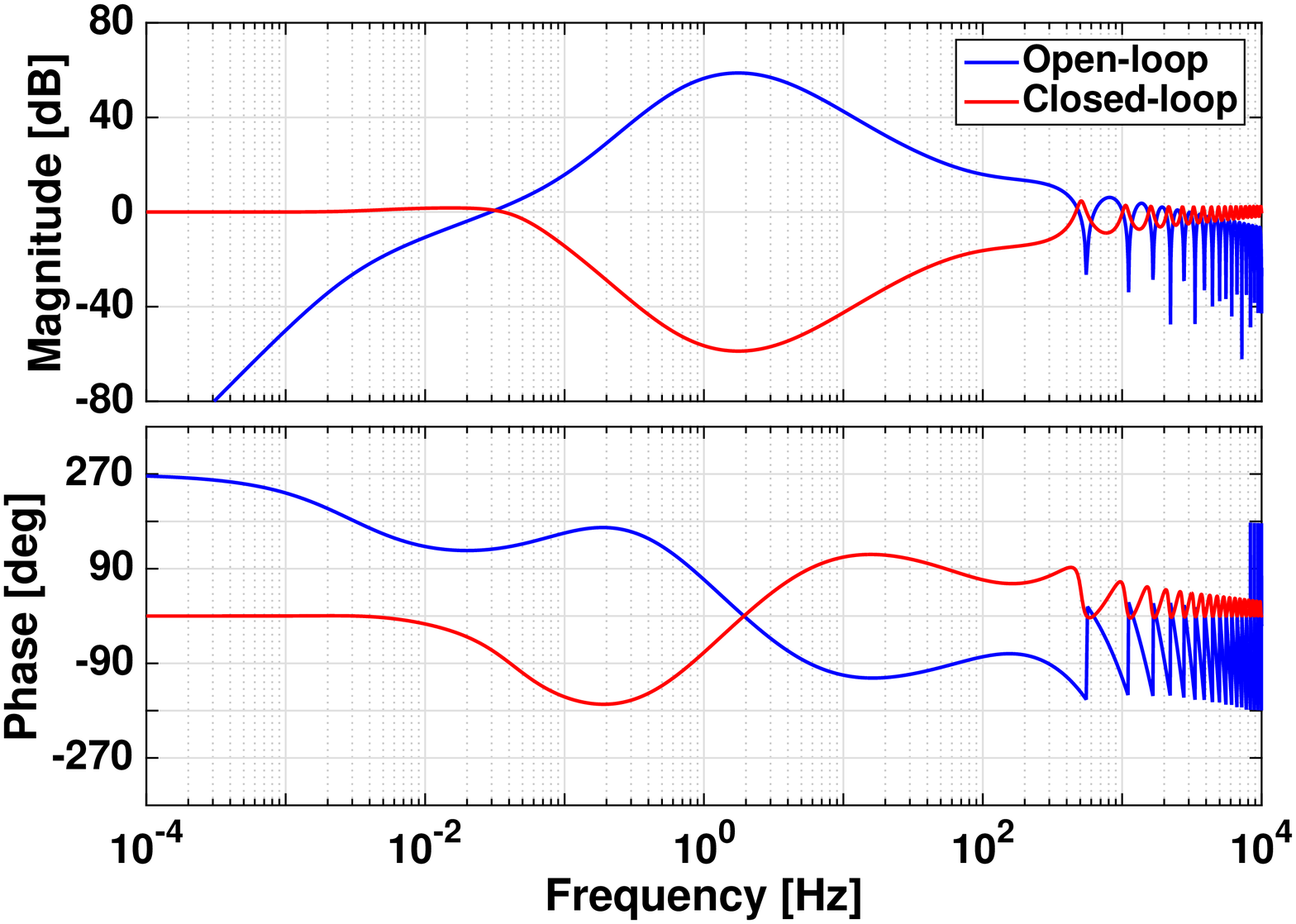}
\caption{Bode plots of open- and closed-loop gain for proposed GRACE-FO arm-locking system.}
\label{fig:bode}
\end{center}
\end{figure}

The goals of the GRACE-FO arm-locking system are to demonstrate strategies that address the two key challenges of arm-locking system design described above: (1) - a controller bandwidth that extends above the arm response null frequencies; and (2) - mitigation of laser frequency pulling through AC coupling to enable stable long-term operation. Figure \ref{fig:bode} shows Bode plots of the open- and closed-loop gains of the proposed control design. The system has a lower unity gain frequency of $\sim 35\,\mbox{mHz}$, which was chosen to keep the expected pulling of the master laser due to time-varying Doppler less than $100\,$MHz, a small fraction of the width of a typical single-mode region in the GRACE-FO lasers. The amount of pulling from the time-varying Doppler was estimated by modifying (\ref{eq:dopPull}) to account for finite open-loop gain and a sinusoidal Doppler shift:

\begin{equation}
\frac{d\nu}{dt} \approx \left(\int_{0}^{\infty} \left|\frac{G}{1+G\left[1-e^{-2\pi i f\tau}\right]}\right|^{2}S_{Dop} df\right)^{1/2} \label{eq:dopPullAC}
\end{equation}
where $G(f)$ is the gain of the arm-locking controller, $\tau$ is the round-trip arm delay, and $S_{Dop}(f)$ is the power spectral density of the time-varying round-trip Doppler signal. The assumption is that the Doppler signal is removed from the arm-locking error signal at lock acquisition, and that that constant offset is maintained for the entire lock period. Using (\ref{eq:dopPullAC}) with the controller design used in Figure \ref{fig:bode} and the estimate of $S_{Dop}(f)$ presented in Section \ref{sec:link} and show in Figure \ref{fig:doppler} yields an estimated Doppler pulling of $\sim74\,$MHz.

After crossing the lower unity-gain frequency at $35\,$mHz, the open-loop gain rises to a peak of $\sim 50\,\mbox{dB}$ at $\sim 2\,\mbox{Hz}$. The gain then drops down with a gradually diminishing power-law index, eventually reaching $G(f)\propto f^{-0.5}$ around $200\,$Hz. The first null of $T_{arm}(f)$ is encountered at $555\,$Hz and approximately four more nulls are bridged before the system passes through its upper unity gain frequency at $\sim 3\,$kHz. The closed-loop gain shows a minimum of $-50\,$dB around $2\,$Hz, meaning that laser frequency noise at those frequencies will be reduced by more than two orders of magnitude. The noise-enhancement features associated with the null frequencies are less than $3\,$dB.

\begin{figure}[h]
\begin{center}
\includegraphics[width=8 cm]{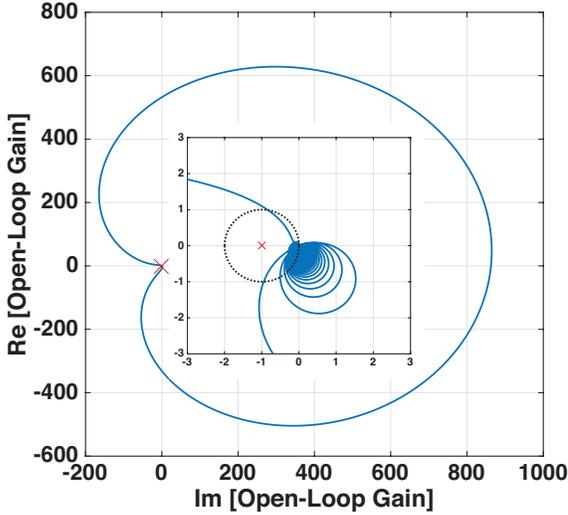}
\caption{Nyquist plot of open-loop gain for proposed GRACE-FO arm-locking system. The inset shows a zoom in of the low-gain region, showing the system's interaction with the noise-enhancement region (interior of the dashed circle).}
\label{fig:nyquist}
\end{center}
\end{figure}

Further detail of the system behavior near the lower and upper unity gain frequencies can be seen in the Nyquist plot of the open-loop gain in Figure \ref{fig:nyquist}. At frequencies well below the measurement band, the open loop gain has negligible amplitude (corresponding to the origin in Figure \ref{fig:nyquist}). As frequencies increase, the system passes through the edge of the noise enhancement region and then rapidly increases in gain as the phase rotates clockwise.  As the system approaches the first null of $T_{arm}(f)$, it crosses into the noise enhancement region and gradually spirals into the origin. The phase margin at the lower unity gain frequency is $54\,$deg whereas the minimum phase margin at the first null is $20\,$deg.

\section{Methodology}
\label{sec:methods}
To validate the proposed arm-locking system design for GRACE-FO a series time-domain numerical models were developed using the \texttt{Simulink\textsuperscript{\textregistered}} software package. These models were based on a model developed for a prior study of arm-locking for LISA\cite{Thorpe11}. Two models were used to produce the results in this paper, a high-fidelity model with a $100\,$kHz sampling rate used to validate the system's stability near and above the upper unity-gain frequency and an economized model with a $4\,$kHz sampling rate used to explore the system's stability at the lower unity gain frequency and its response to Doppler shifts caused by the GRACE-FO orbits. The two models were cross-checked in an overlapping frequency regime and found to be consistent with one another (see section \ref{sec:results} for further details).

In the high-fidelity simulation, the GRACE-FO model is divided into three main components: the master spacecraft, the repeater spacecraft, and a two-way link model. The primary state variable in each simulation is frequency, which at various points in the simulation represents optical frequencies (relative to some fixed reference frequency), frequencies of electrical signals, or frequencies of digital signals. 

\subsection{Spacecraft model}
\label{sec:SC}
Each spacecraft model consists of a laser model, an interferometer model, and a controller. In the laser model, a random noise stream is generated and filtered to reproduce the expected `free-running' frequency noise of the LRI lasers, $\delta\tilde{\nu}(f)\approx 40\,\mbox{kHz}/\mbox{Hz}^{1/2}\cdot (f/1\,\mbox{Hz})^{-1}$.  A frequency correction command is added to the intrinsic laser noise to produce the laser output. This command is filtered with a simple pole at $100\,$kHz which represents the frequency response of the piezo-electric frequency actuator on the LRI laser system. This single-actuator model is a simplification of the actual laser system, which includes both a high-bandwidth, low-dynamic-range piezo actuator and a low-bandwidth, high-dynamic-range thermal actuator.  

The interference of the outgoing and incoming optical beams is modeled as a simple subtraction of the incoming and outgoing frequencies plus an additive noise.  This represents the optical interference, conversion to electrical signal in the photoreceiver, and extraction of the interference phase (or frequency) time-series by the Laser Ranging Processor (LRP). The additive noise is modeled as the shot noise associated with making a frequency measurement of $1064\,\mbox{nm}$ light with $\sim100\,$pW of received power. As shown in section II.F.2 of \cite{Thorpe11}, this has an equivalent frequency noise of $\delta\tilde{\nu}_{shot}= 43\,\mu\mbox{Hz}/\mbox{Hz}^{1/2}\cdot(f/1\,\mbox{Hz})$.

The controller model represents the digital filter applied by the LRP to convert the measured interference phase into a command for the laser frequency actuators. The transfer function for this filter differs between the master and repeater spacecraft.  For the repeater, the filter has a simple $f^{-1}$ transfer function with a unity gain frequency of $35\,$kHz.  For the master, the arm-locking controller is implemented in several stages with an overall transfer function,

\begin{equation}
G_{m}(f) = G_0\cdot G_1^2\cdot\left[G_2 + G_3\right]\cdot G_4\cdot G_5^2,
\label{eq:Gmaster}
\end{equation}
where $G_i$ represent the transfer functions of the individual stages. 

Table \ref{tab:controller} summarizes the content and function of each stage.  Stage 0 is a constant gain stage used to set the upper unity gain frequency of the arm-locking system. The system response plotted in Figure \ref{fig:bode} and described in section \ref{sec:results} used a gain of 3. Stage 1 consists of two substages arranged in series, with each substage having two pole-zero pairs. The function of Stage 1 is to boost the system gain in the `measurement' band around $1\,$Hz. Stage 2 is a transition stage to reduce the gain from its max value around $1\,$Hz as Fourier frequency increases. Stage 3 is a composite of nine single-pole filters arranged in parallel that results in a transfer function with an equivalent response of $G_3(f)\propto f^{-0.5}$. This provides the additional phase margin necessary to maintain stability around the nulls in $T_{arm}(f)$. Stages 4 and 5 combine to reduce the system gain at low frequencies and avoid frequency pulling. Stage 5 is two simple differentiators in series while stage 4 consists of two poles at $3\,$mHz to roll off the $f^2$ response of Stage 5.

\begin{table}[h!]
\centering
\caption{\label{tab:controller}Summary of arm-locking controller subcomponents. The overall controller is built from these components as described in Equation \ref{eq:Gmaster} and the accompanying text. }
\begin{tabular}{|c|c|c|}
\hline
Stage & Description & Function\\
\hline
0 & $k = 3$ & gain stage \\
\hline
1 & 
{$\!\begin{aligned}
               k &= 1 \\    
               p &= 70\,\mbox{mHz},\: 3\,\mbox{Hz} \\
               z &= 100\,\mbox{mHz},\: 100\,\mbox{Hz} \end{aligned}$} 
& in-band gain \\
\hline
2 & 
{$\!\begin{aligned}
               k &= 251 \\    
               p &= 1\,\mbox{Hz} \end{aligned}$} 
& transition stage \\
\hline
3 & 
{$\!\begin{aligned}
               k =& (99,\:147,\:218,\ldots \\
               & 323,\:480,713,\ldots \\
                &1.06e3,\:1.57e3,\ldots \\
                &2.33e3) \\    
               p = &(50,\:105,\:221,\ldots \\
               &463,\:972, 2.04e3,\ldots \\
               &4.29e3,\:9.01e3,\ldots \\
               &1.89e4)\,\mbox{Hz}\end{aligned}$} 
& cross upper UGF \\
\hline
4 & 
{$\!\begin{aligned}
               k &= 1 \\    
               p &= 3\,\mbox{mHz},\: 3\,\mbox{mHz} \end{aligned}$} 
& roll-off AC coupling \\
\hline
5 & differentiator ($z=0\,\mbox{Hz}$) & AC coupling\\
\hline
\end{tabular}
\end{table}%

\subsection{Link model}
\label{sec:link}
\begin{figure}[h]
\begin{center}
\includegraphics[width=8 cm]{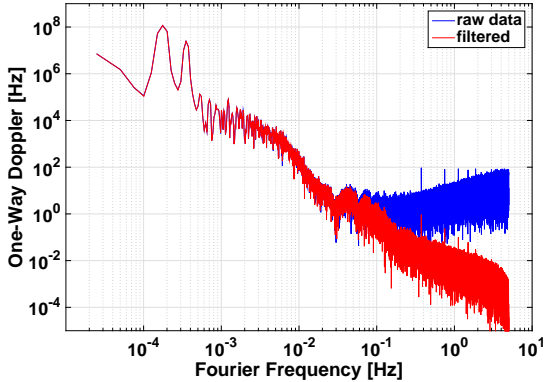}
\caption{Spectra of representative Doppler shifts for the GRACE-FO LRI derived from microwave ranging data from GRACE. The blue curve shows the raw GRACE microwave ranging data\cite{Tapley04, Dunn2003, Case2010} after finite differencing and scaling to Doppler shift for $1064\,$nm light. The red curve, which corresponds to the signal used in the simulations, has been low-pass filtered to reject excess high-frequency noise that originates in the GRACE microwave system.}
\label{fig:doppler}
\end{center}
\end{figure}

The two primary functions of the link model are to model the propagation delay between the two spacecraft, which is approximately $1.8\,$ms and varies by $\sim 7\,\mu$s over a full orbit. This variation is less than the resolution of the 100kHz sampling rate for the model, so the propagation delay is modeled using a simple buffer. The second component of the link model is the Doppler shift that results from the relative motion between the two spacecraft. The relative motion between the spacecraft is caused by differences in the spacecraft orbits, anomalies in the geoid (the primary science signal for GRACE and GRACE-FO), and uncorrelated atmospheric drag in each spacecraft. The blue curve in Figure \ref{fig:doppler} shows a spectrum of Doppler shifts that were derived from existing data taken from the microwave ranging instrument on GRACE. To convert the GRACE ranging data to equivalent Doppler shift, the range signal is finite-differenced to generate the range rate in units of m/s and then scaled by $\lambda^{-1}$, where $\lambda=1064\,$nm is the wavelength of the LRI laser light. This conversion procedure is appropriate for the contribution to the GRACE-measured range from true spacecraft motion. However, the GRACE data also includes a white noise floor at approximately $1\,\mu m/\mbox{Hz}^{1/2}$ which is converted to an equivalent Doppler noise of $\sim 1\,\mbox{Hz}/\mbox{Hz}^{1/2}\cdot(f/1\,\mbox{Hz})$. This Doppler noise exceeds the contribution to the Doppler from true ranging for Fourier frequencies greater than $\sim 300\,\mbox{mHz}$. To avoid introducing this non-physical excess Doppler noise into the link model, we filter the derived Doppler signal with a 128-point Bartlett-windowed FIR lowpass filter operating at the 10Hz sampling frequency of the original GRACE data. The spectrum of this filtered Doppler signal is shown in the red trace in Figure \ref{fig:doppler}. For frequencies above 10Hz, the Doppler signal is linearly interpolated.

\subsection{Economized model}
The $100\,$kHz sampling rate of the high-fidelity model makes it computationally expensive to run long-duration simulations; on a standard laptop, the high-fidelity model runs more than 30x slower than real-time. This sampling rate is necessary to sufficiently resolve the system behavior near the upper unity gain frequency of $3\,$kHz. To facilitate longer simulations necessary to study the frequency pulling, we constructed a economized model with a sampling rate of $4\,$kHz. In the economized model, the two-way link model and the repeater spacecraft model were replaced with a single round-trip link model. This model approximates $T_{arm}(f)$ as a first-order derivative plus two poles at $300\,$Hz and a zero at $1\,$kHz. The Doppler frequency is doubled to account for the round-trip.  The master spacecraft controller is modified by removing $G_3(f)$ and allowing $G_2(f)$ to carry the system through unity gain at $200\,$Hz.  Below $\sim10\,$Hz, the response of the economized and hi-fidelity controller are identical, making the economized model an appropriate tool to study low-frequency behavior.

\section{Results}
\label{sec:results}
Three simulations using the models described in section \ref{sec:methods} were conducted to validate the GRACE-FO design from section \ref{sec:design}. For each simulation, the random seeds used to generate noise for the lasers and shot noise were initialized to the same value and the same portion of filtered Doppler data from Figure \ref{fig:doppler} was used. To initialize the simulation, the master controller was left open for $5\,$ms to allow the laser noise to propagate through the arm. The instantaneous value of the round-trip Doppler was then subtracted from the error signal and the master controller was enabled.  The Doppler offset was held fixed for the duration of the simulation. In actual practice, an estimate of the round-trip Doppler would be made either by averaging the science signal as described in \cite{McKenzie_09} or by dead-reckoning from orbital ephemerides and GPS data. 

Two of the simulations were conducted with the high-fidelity model, one with a duration of $10\,$s and output data saved at the full $100\,$kHz rate and the other with a duration of $500\,$s and the output data filtered and downsampled to $10\,$Hz.  The third simulation was made using the economized model with a duration of $10\,$ks and output data filtered and downsampled to $10\,$Hz. 

\subsection{Lock Acquisition and frequency pulling}
\begin{figure}[h!]
\begin{center}
\includegraphics[width=8 cm]{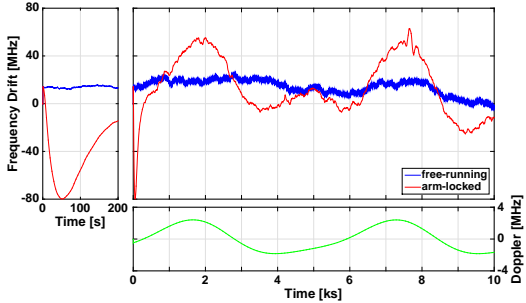}
\caption{Timeseries of master laser frequency and round-trip Doppler shift from a $10\,$ks simulation with the economized model. The top panels compare the drift from nominal frequency for a free-running laser (blue) and laser locked to the GRACE-FO arm (red). After an initial lock acquisition transient lasting a few hundred seconds (top left) the arm-locked system undergoes roughly periodic drifts with an amplitude of $\sim 35\,$MHz, well within an acceptable operating range for the GRACE-FO laser. The bottom panel shows the round-trip Doppler shift, which is also roughly equivalent to the drift in the optical beat note at the master spacecraft. The drift of the arm-locked laser is correlated with the Doppler shift at low frequencies due to the frequency-pulling effect described in the text.}
\label{fig:pulling}
\end{center}
\end{figure}

Figure \ref{fig:pulling} shows a timeseries of the master laser frequency for the $10\,$ks simulation with the economized model in red.  The drift of the free-running laser is shown in blue for comparison.  After an initial lock-acquisiton transient of $\sim 80\,$MHz, the arm-locked laser frequency exhibits oscillations around the free-running noise with an amplitude of $\sim 35\,$MHz and a period of $90\,\textrm{min}\sim5.4\,$ks.  These oscillations are correlated with the round-trip Doppler signal during this period, which is shown in the lower panel of Figure \ref{fig:pulling}.

Overall, a drift in the master laser frequency of less than $100\,$MHz should be easily tolerated by the LRI laser subsystem. Typical lasers of this type have single-mode operation regions that are several GHz wide.  The optical beat note at the repeater spacecraft is maintained at a constant offset by the high-gain phase-lock loop. The drift in the optical beat note at the master spacecraft is essentially equal to the round-trip Doppler shift, with a peak-to-peak amplitude of about $4\,$MHz, less than the $\pm6\,$MHz allowed by the LRI signal chain.

\subsection{System stability and performance}
\begin{figure}[h]
\begin{center}
\includegraphics[width=8 cm]{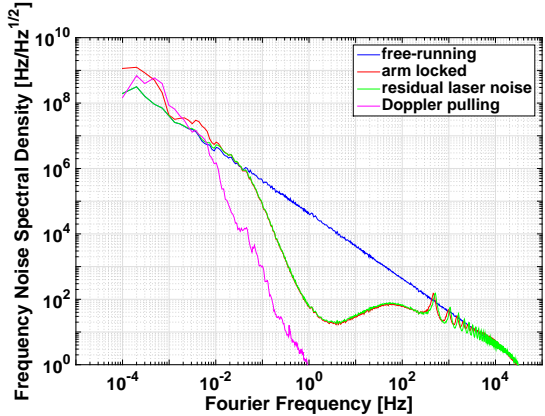}
\caption{Amplitude spectral density of frequency noise in a simulated GRACE-FO arm-locking system. The free-running frequency noise is shown in blue and the arm-locked frequency noise is shown in red. The green trace shows the expected residual frequency noise based on the system transfer function (gain limit) while the magenta trace shows the noise associated with pulling of the laser frequency caused by the time-varying Doppler signal. All traces derived from two simulations with the high-fidelity model: a $10\,$s simulation with $100\,$kHz data sampling and a $500\,$s simulation with $10\,$Hz data sampling.}
\label{fig:noise}
\end{center}
\end{figure}

Figure \ref{fig:noise} shows the amplitude spectral density of frequency noise from the three simulations described above. The blue trace shows the free-running noise of the master laser, which has an approximate noise spectral density of $40\,\mbox{kHz}/\mbox{Hz}^{1/2}\cdot(f/1\,\mbox{Hz})^{-1}$. The red trace shows the noise of the arm-locked laser, which is less than $100\,\mbox{Hz}/\mbox{Hz}^{1/2}$ in the band $1\,\mbox{Hz} < f < 100\,\mbox{Hz}$. The green trace shows the expected residual frequency noise resulting from the finite-gain of the arm-locking loop. It is computed by multiplying the closed-loop gain of the arm-locking system (red trace in Figure \ref{fig:bode}) by the free-running noise spectral density (blue trace in Figure \ref{fig:noise}). The fact that the green trace closely matches the red trace confirms that the system is gain-limited over the entire active bandwidth. The next largest contribution, the pulling from the Doppler noise, computed using (\ref{eq:dopPullAC}) and shown in magenta. The Doppler noise contribution has a spectral density of $\sim 1\,\mbox{Hz}/\mbox{Hz}^{1/2}\cdot(f/1\,\mbox{Hz})^{-2}$, which is more than two orders of magnitude below the residual frequency noise from the band extending from $1\,$Hz down to the lower unity-gain frequency of $\sim 35\,$mHz. 

\section{Discussion}
\label{sec:discuss}
The simulation results in section \ref{sec:results} demonstrate that it is possible to design an arm-locking system for the GRACE-FO LRI that achieves an interesting level of frequency stability while maintaining pulling of the master laser frequency within acceptable limits.  The frequency noise performance is at a similar level as that of the primary LRI stabilization, an optical-cavity based system with a performance of $34\,\mbox{Hz}/\mbox{Hz}^{1/2}$. This is with the significant caveat that the best arm-locking performance is at $1\sim10\,$Hz, a higher Fourier frequency than the cavity, which meets its performance goal in the GRACE-FO LRI band of $2\,\mbox{mHz} < f < 1\,\mbox{Hz}$.

It would be possible to extend the bandwidth of the GRACE-FO system to somewhat lower frequencies by reducing the lower unity-gain frequency. However, this would come at the expense of increased laser frequency pulling and would eventually result in the system performance being limited by pulling from in-band Doppler noise. For LISA, the much larger separation in frequency between the science band and the primary Doppler modulation frequencies, coupled with the much simpler Doppler signal (no geoid signal) means that the trade between frequency pulling and low-frequency gain is not as tightly constrained. The LISA arm-locking designs presented in \cite{McKenzie_09} and \cite{Thorpe11} demonstrate how this trade can be addressed for LISA.

While existing analytic, numerical, and experimental work demonstrating various aspects of arm-locking for LISA has significantly reduced the technical risk for this technique, an
on-orbit demonstration with GRACE-FO would provide further confidence that this technique can be implemented in a real-world environment.  Such a demonstration could be conducted with no modification to the LRI hardware; a software update would provide the necessary functionality to conduct such an investigation. 

The demonstration of arm-locking on GRACE-FO could influence the choice frequency stabilization system for LISA, which is currently baselined as an external optical cavity.  Arm-locking can dramatically benefit an interferometry system for LISA, either by replacing the optical cavity and reducing mass, power, and cost or by augmenting the cavity and providing increased frequency stability that can be used to trade other system design requirements. An on-orbit demonstration of arm-locking with GRACE-FO would validate arm-locking as a technique and lower the cost and risk of including it in LISA.

More generally, a demonstration of arm-locking, combined with the previously-mentioned demonstration of TDI using GRACE-FO\cite{Francis15}, would provide an on-orbit demonstration of several of the key technologies for long-baseline interferometry in LISA. Such a demonstration would be an excellent complement to the soon-to-be-launched LISA Pathfinder mission that demonstrates force disturbance reduction through precision drag-free control and short-baseline interferometry of freely-falling test masses\cite{LPF_LISAX}. The combined results of both efforts would place LISA technologies on an extremely firm foundation, lowering the barriers for implementing this exciting mission.

\par

\begin{acknowledgements}
The authors would like to acknowledge Andrew Sutton for his thoughtful commentary during the early stages of this work and Shannon Sankar for a thorough review of the final manuscript.
\par
This work was supported by the NASA Research Opportunities in Space and Earth Sciences (ROSES) program under the grant 11-APRA11-0029. Part of this research was performed at the Jet Propulsion Laboratory, California Institute of Technology, under contract with the National Aeronautics and Space Administration (NASA).
\par
Copyright (c) 2015 United States Government as represented by the Administrator of the National Aeronautics and Space Administration. No copyright is claimed in the United States under Title 17, U.S. Code. All other rights reserved.
\end{acknowledgements}

\bibliography{Bibliography.bib}
\end{document}